# The Magic of Slow-to-Fast and Constant: Evaluating Time Perception of Progress Bars by Bayesian Model

Qihan Wang[1], Xinyue Kang[2], Pei-Luen Patrick Rau[2]

[1]Yuanpei College, Peking University, Beijing, China

[2] Department of Industrial Engineering, Tsinghua University, Beijing, China

**Corresponding author:** Pei-Luen Patrick Rau; Email: rpl@mail.tsinghua.edu.cn; Phone No.: +86 010-62776664

**Running head:** EVALUATING TIME PERCEPTION OF PROGRESS BARS

**Manuscript type:** Research Article

**Word count:** 4239

**Funding Sources:** This work was funded by National Natural Science Foundation of China 71942005.

## Abstract

**Objective**: We aimed to use adaptive psychophysics methods, which is a Bayesian Model, to measure users' time perception of various progress bar quantitatively.

**Background**: Progress bar informs users about the status of ongoing processes. Progress bars frequently display nonuniform speed patterns, such as acceleration and deceleration. However, which progress bar is perceived faster remain unclear.

**Methods**: We measured the point of subject equality (PSE) of the constant progress bar toward four different 5-second progress bars with a non-constant speed. To measure PSE, in each trial, a constant progress bar and a non-constant progress bar were presented to participants. Participants needed to judge which one is shorter. Based on their choice, the model generated the time duration of constant progress bar in next trial. After 40 trials for each non-constant progress bar, the PSE was calculated by the model. Eye tracking was recorded during the experiment.

**Results**: Our results show that the constant progress bar and speed-up progress bar are perceived to be faster. The anchoring effect fits the results of our study, indicating that the final part of the progress bar is more important for time perception. Moreover, the eye-tracking results indicate that the progress bar is perceived to be slower is related to the overload of cognitive resources.

**Conclusion**: The constant progress bar and speed-up progress bar are perceived as the quickest.

**Application**: The results suggest that UX design can use constant or speed-up progress bar, in order to improve user experience in waiting.

**Précis:** This study used the Bayesian adaptive models called QUEST to measure the point of subject equality of different progress bars towards constant progress bar. The results were analyzed to calculate which progress bar is perceived as shorter. To explain the results, eye-tracking data were recorded during the experiment.

# The Magic of Slow-to-Fast and Constant: Evaluating Time Perception of Progress Bars by Bayesian Model

The progress bar is a graphical control element used to visualize the progress of extended computer operation. It informs users about the status of ongoing processes such as loading an app, submitting a form, or uploading a document. The linear progress indicator is a widely used progress bar. There are two different linear progress bars: determinate indicators, which display the indicator increasing in width from 0 to 100% of the track, in sync with the progress of the process, and indeterminate indicators, which express an unspecified amount of wait time. Many progress bars used in UX design are linear determinate indicators, such as webpage loading in Safari and advertisements on YouTube. People judge how long they must wait based on the motion of the progress bar. Progress bars frequently display nonuniform speed patterns, such as acceleration and deceleration (Harrison et al., 2007).

Although the waiting state on smartphone interfaces is unavoidable when using interactive applications (Li & Chen, 2019), waiting is not easy for humans. Individuals tend to overestimate wait duration (Hornik, 1984). Even if they can wait, they may not want to wait if they believe the loading time is excessively long (Roberts & Fishbach, 2022). If a person watching a YouTube advertisement thinks the progress bar lasts too long, they may quit waiting and close the website. The website may significantly lose advertisement income and the number of users because of people's poor patience in waiting. As the progress bar always appears in the waiting scene as an indicator of how long users should wait, studying how to make the progress bar appear faster is essential for the user interface.

The human perception of time duration is not as accurate as that of the clock. Many physical properties influence human cognition over time. In the physical world, speed is an essential and pervasive variable in daily life. Some studies have proven that the human perception of duration is affected by stimulus speed and speed changes (Brown, 1995; Matthews, 2011). Although it is unclear what speed changes make people feel faster, an increase in speed can lead to overestimations of durations and vice versa (Karşılar et al., 2018). This is because the way speed affects time duration is modulated by many other factors, such as its size (Thomas & Cantor, 1976) and brightness (Eagleman & Paryadath, 2009). Hence, speed changes may influence the time perception of progress bars. Therefore, it is important to carefully investigate how speed changes affect how people perceive the duration of progress bars.

## Related Work

Although some previous studies have investigated the differences in human time perception for progress bars with different speed changes, the results have varied significantly.

Some studies suggest that slow-to-fast progress bar is perceived to be shorter. Harrison et al. (2007) studied nine progress bars, claiming that users are most willing to tolerate negative progress behavior at the beginning of an operation. Hence, slow-to-fast progress bar had the best effect on tolerance. Additionally, users pay more attention to the end position of visual feedback (Yokoyama et al., 2021). Because of this, users prefer slow-to-fast progress bar.

Conversely, some studies suggest that the “fast-to-slow” progress bar is perceived as shorter. Conrad et al. (2010) studied the relationship between the speed of progress bars and task completion rate, showing that when early feedback indicates fast progress, participants are less likely to abandon it. Another meta-analysis study also showed that compared to a constant progress indicator, people have more patience for fast-to-slow indicators and less patience for slow-to-fast indicators (Villar et al., 2013).

Some studies demonstrate the advantages of a constant progress bar. Branaghan and Sanchez (2009) suggested that the constant progress bar performed at the top of both reasonableness and preference, keeping users informed without increasing arousal or focusing attention on temporal stimuli. Another study showed that participants preferred a progress bar that moved consistently (Amer & Johnson, 2014).

The current results of these studies cannot reach an agreement on which progress bar is perceived as shorter, probably because the methods they used are not precise enough to tell the difference; most of the papers mentioned above simply ask the participants to watch the progress bar once and then let the participants estimate the wait duration of the progress bar (Li & Chen, 2019), choose the progress bar they like best or rate the progress bar through a 7-point Likert scale (Rose et al., 2019). Differences in time perception can be minimal and difficult to detect. Moreover, seeing each progress bar only once is unreasonable because the human perception of time may differ slightly from trial to trial. Hence, these methods may contain some noise. Therefore, a more accurate and precise method is required to solve this problem.

Most previous studies do not provide clear explanations for their results, for example, why human cognition tends to think some progress bars move shorter than others. Some researchers have mentioned the arousal of focusing attention (Branaghan & Sanchez, 2009) and people’s selective attention toward the end position of visual feedback (Yokoyama et al., 2021).

## Current Work

In our study, we used psychophysics to measure the time perception of progress bars. Psychophysics is widely used to study the perceptual system quantitatively, utilizing quantitative methods to examine the relationship between physical stimuli and the perception they elicit (Green et al., 1966). In psychophysics, the point of subject equality (PSE) is defined as any point along a stimulus dimension at which a variable stimulus (visual, tactile, auditory, etc.) is judged by an observer to be equal to a standard stimulus (Vidotto et al., 2019). To test the PSE, in each trial, participants were asked to see one variable stimulus and the standard stimulus and respond to the demand question ("which progress bar is shorter," etc.). By analyzing all the trial results, the variable stimuli with a 50% probability of a particular psychophysical response are the PSE. In this study, we measured the PSE of different progress bars towards the constant progress bar so that which progress bar is perceived as shorter can be detected more accurately.

Compared to the classic methods of psychophysics that use preset variable stimuli, adaptive procedures that use previous responses to guide further testing are much more effective (Simpson, 1989). In our study, a Bayesian adaptive psychometric method called QUEST was used to estimate each participant's point of subject equality of the progress bar with constant speed to progress bars with non-constant speeds. QUEST's rationale is to fit a Weibull function to the incoming data to model the psychophysical transformation between the stimulus intensity and sensation magnitude (Watson & Pelli, 1983). The positioning of the parameter of interest along the intensity dimension was approximated using Bayesian maximum likelihood estimation. Given a sensible prior assumption for such a distribution, the algorithm determines the most informative intensity to which the participant should respond (Taesler & Rose, 2017). In other words, in our study, the QUEST procedure parametrically calculated the variable stimulus of the subsequent trials (i.e., the duration of the progress bar) based on participants' decisions on previous trials and consequently calculated the point of subject equality for the constant progress bar toward non-constant progress bars. For instance, the length of a constant-speed progress bar is considered to be as long as a five-second acceleration progress bar. A previous study also mentioned that adaptive experiments for progress bars should be conducted because this would allow the quantitative evaluation of relative perceptual variations (Harrison et al., 2007). To the best of our knowledge, this is the first time that the Bayesian psychophysical approach has been used to measure human time perception in human-computer interactions.

The explanation for why human perceived some progress bars as shorter is still unclear. In our study, eye tracking was used to record the participants' eye movements when the progress bar moved. Previous studies have shown that the eye blink rate (EBR) is related to many cognitive functions. Since EBR is related to dopamine (DA) activity, it predicts many DA-related cognitive performances (Jongkees & Colzato, 2016): cognitive flexibility, attention, and working memory updating (Dang et al., 2016; Zhang et al., 2015). Hence, to explain the results, we aimed to use eye tracking to investigate how cognitive performance influences human perception of time.

In this study, we used the Bayesian adaptive psychophysics method QUEST to measure the PSE of different progress bars towards constant progress bar. The results were analyzed to calculate which progress bar is perceived as shorter. To explain the results, eye-tracking data were recorded during the experiment.

## Methods

### Participants

Twenty participants were invited to participate in our study (11 females and 9 males). Their ages ranged from 18–29 years old (*M*=23.9 years). All participants were students from Tsinghua University. All participants had experience using computers to watch progress bars. This study was approved by the Department of Industrial Engineering of Tsinghua University.

**Materials**

Four nonlinear functions were developed to embody progress bars with different speed changes: the Bezier progress bar, slow-down progress bar, speed-up progress bar, and elasticity progress bar. A progress bar with constant speed was included as the baseline for comparison. Table 1 and Figure 1 describe the behavior of each progress function.

The stimuli were presented using PsychToolbox in MATLAB (MathWorks, Ltd.). The experiment was conducted on a 23” display screen running at 1920 × 1080. An external USB keyboard was attached to the laptop so participants could key their responses. The length of the progress bar is 600 × 20 pixels. The time duration of the nonconstant progress bar was 5s. The duration of the constant progress bar changes in each trial, which is calculated using the Bayesian model. The progress bar is displayed at the center of the screen with a white background. A central fixation point with a size of 2 × 2 pixels was presented before each trial. The refresh frequency was set to 120 Hz.

The 7Invensun aSee Glasses Eye Tracker (120 Hz) was used to record eye tracking. The EBR was defined as the ratio between the total number of blinks in a phase and the duration of the phase in seconds.

**Table 1**

*The Experimental Progress Functions*

| Name | Description | Rate Trend | Acceleration | Function |
|---|---|---|---|---|
| Constant | Progresses linearly | Constant | None | f(x)=x |
| Bezier | Smooth curve | Speeds up, then Slows down | Non-Constant | shown below |
| Slow Down | Decelerates | Slows down | Constant | $f(x)=24x^2$ |
| Speed Up | Accelerates | Speeds up | Constant | $f(x)=240x-24x^2$ |
| Elasticity | Elastic impact | Speeds up | Non-Constant | f(x)=e^(-β*x)*(A-β*A*x)-A; β=-1/5, A=600*e/(2-e) |

*Note.* x refers to time*Flip Interval (120), for example, if time=1s, x=120.

Bezier function is defined as $\sum_{N}^{5}{}_{=1}$ (1-x/600)^(N-k-1)*p(k+1,2)*(x/600)^k*t(N,k+1).

**Figure 1**

*Graphs of Progress Functions*

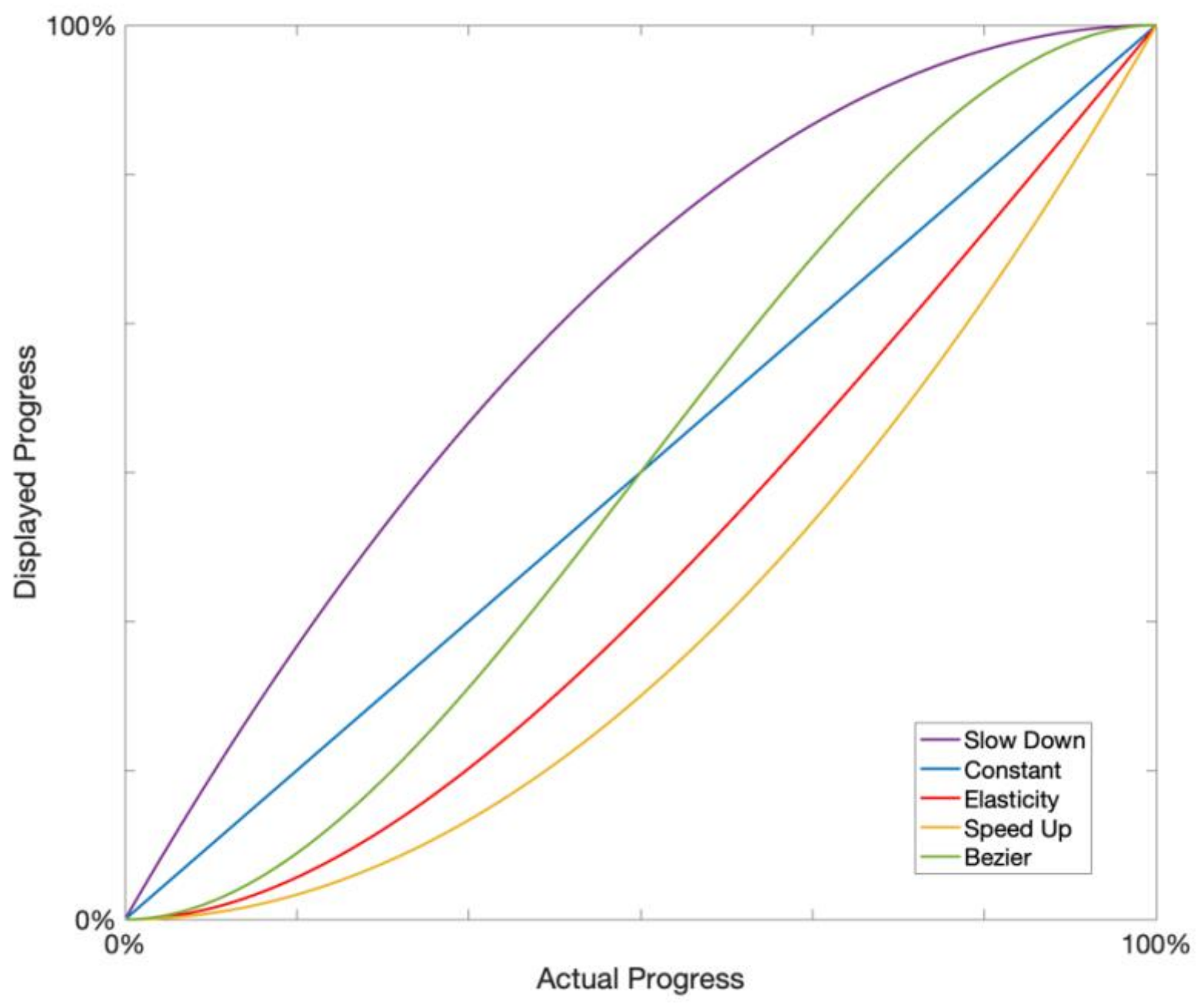

## Procedure

Before the experiment started, the participants were given a brief introduction and then asked to wear the eye-tracking device with the experimenter's help. The participants were then told to choose a shorter progress bar for each trial. Subsequently, three exercise trials with feedback were presented to the participants. In the formal experiment, participants did not receive any feedback.

This study measures the PSE of the constant progress bar towards four nonlinear function progress bars. To measure PSE, 40 trials were presented to each participant. In each trial, the fixation point was first presented for 2s, and then the standard stimulus (which refers to the non-linear function progress bar that lasts for 5s) and the variation stimulus (which refers to the linear function progress bar whose duration is updated and calculated by the calculation of QUEST for each trial) were presented. After the presentation, the instruction prompted the participants to judge which progress bar had a shorter time by pressing the keyboard. As four PSE were measured, there were 160 trials in total. After every 40 trials, participants were allowed to rest for one minute. The total task time was approximately 40 minutes. The order of the presentation was counterbalanced in two ways. First, the order of standard stimulus and variation stimulus were random for each trial. Second, the order of measurement of the PSE of the four progress bars was also random.

## Data Analysis

### *PSE Measurement*

The standard implementation of the QUEST procedure was provided by the Quest toolbox in Psychtoolbox Version 3, which is a Bayesian toolbox for testing observers and estimating their thresholds. Different algorithms can be used to calculate the best intensity for subsequent trials.

Here, we use the algorithm of QuestMean, which is each trial, and the final estimate is at the MEAN of the posterior PDF to suggest the best intensity for each trial.

### *PSE Differences Across Groups*

To test whether the PSE of different progress bars was different in the four curves, repeated measures ANOVA was performed with the within-subject factor curve.

### *Differences Between Non-constant Speed and Constant Progress Bar*

To test whether the PSE of different progress bars was different between the non-constant speed progress bar and constant progress bar, multiple one-sample t-tests were performed.

### *Eye Tracking Data Analysis*

Spearman regression was used to determine whether the PSE of different progress bars correlated with the EBR. Because the EBR has a large inter-individual variability (Magliacano et al., 2020), we generated the rank of the PSE and the EBR of different curves for each participant and used these data to compute the Spearman regression.

## Results

### Point of Subject Equality of Different Progress Bar

The mean values and standard deviations for the PSE of each progress bar are presented in Table 2. The PSE indicates how long the time duration of the constant progress bar is equivalent to that of the non-constant progress bar in 5s. For example, the PSE of Bezier is 5.238s, which means that the constant progress bar in 5.238s is perceived as equal to the Bezier progress bar in 5s, emphasizing that the Bezier progress bar is perceived to be longer than the constant progress bar. The results show that all curves are perceived to be slower than the constant progress bar. Among the four progress bars, the average PSE from largest to smallest is slow-down, Bezier, elastic, and speed-up.

**Table 2**

*Average PSE of Four Different Curves*

| Progress Bar | Point of subject equality(s) M(SD) |
|---|---|
| Bezier | 5.238 (0.370) |
| Speed Up | 5.104 (0.171) |
| Slow Down | 5.672 (0.746) |
| Elasticity | 5.232 (0.234) |

*Note.* The PSE is defined as the time duration of the constant progress bar, which is perceived as equivalent to the non-constant progress bar.

## Participants' Preference Toward Different Progress Bar

Repeated-measures ANOVA revealed significant within-subject effects. $F(3,57)=6.6571$, $p=0.0006$. The mean values and standard deviations for the PSE of each progress bar were presented in Table 2. Multiple comparison results showed that the slow-down progress bar's PSE was significantly higher than the Bezier progress bar ($MD=0.433$, $p=0.014$), elasticity progress bar ($MD=0.440$, $p=0.012$), and speed-up progress bar ($MD=0.568$, $p<0.001$). No significant differences were detected among the other progress bars. Figure 2 shows the progress bar's mean values and standard errors for each PSE.

**Figure 2**

*PSE of Different Curves*

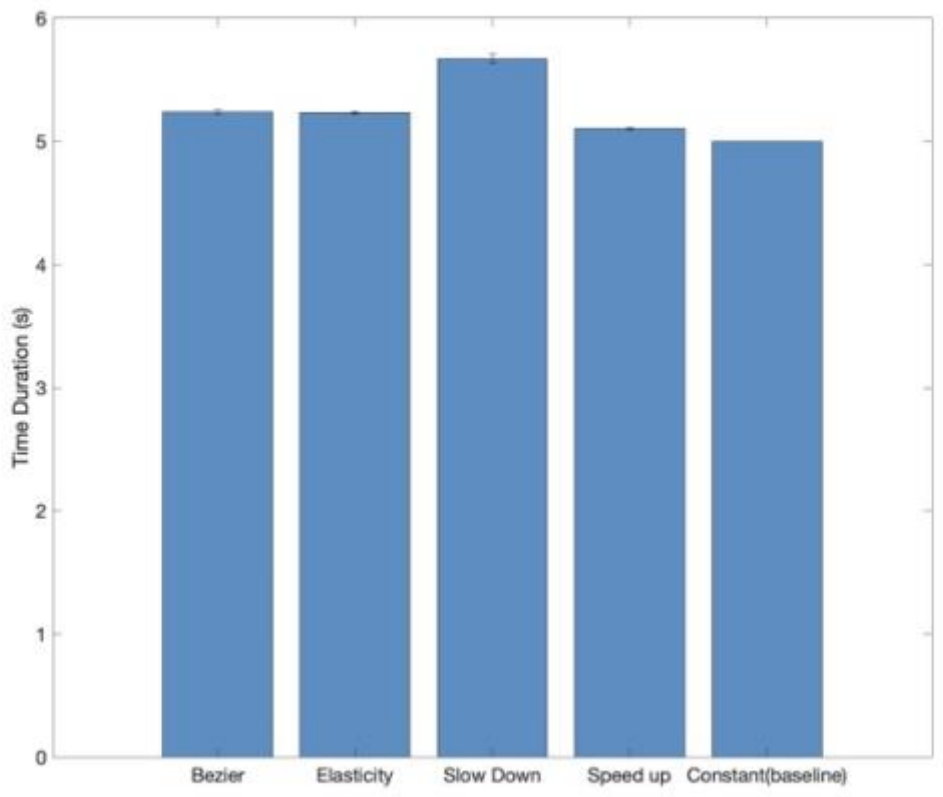

Multiple one-sample t-tests showed that the duration of the constant progress bar was perceived to be significantly shorter than the Bezier progress bar ($t$=2.420, $p$=0.018), elasticity progress bar ($t$=2.356, $p$=0.021), and slow-down progress bar ($t$=6.816, $p$<0.001). There was no difference between the constant progress bar and the speed-up progress bar ($t$=1.052, $p$=0.296).

**Eye Tracking Result**

Statistical tests with Spearman regression analyses showed a significant negative correlation between the PSE of the progress bars and blink rates ($r$=-0.271, $p$=0.012). This indicates that the lower the PSE, the higher the eyeblink rate.

## Discussion

In general, we used the Bayesian model to measure the PSE of the constant progress bar to the non-constant progress bars, which lasted for 5s. The results show that all the non-constant speed progress bars are perceived as slower than the progress bar at a constant speed. Among the four progress bars, the order of the average PSE from largest to smallest is as follows: slow-down, Bezier, elasticity, and speed-up. In other words, the time perception from slowest to fastest is slow-down, Bezier, elastic, and speed-up. The statistical results show that the slow-down, Bezier, and elasticity progress bars are significantly slower than the progress bars with constant speed, while there is no significant difference between speed-up and constant progress bars. Moreover, Bezier, elasticity, and speed-up progress bars are perceived as faster than the slow-down.

Our findings show that the constant speed progress bar is perceived as the fastest, consistent with the most influential model of time perception—the internal clock model (Wang & Wöllner, 2019). This model suggests that humans possess an inner pacemaker that keeps track of time through the accumulation of pulses (Matthews et al., 2011). In other words, the internal pacemaker or oscillator emits pulses that are accumulated until the stimulus offset, at which point the number of counts in the accumulator provides a measure of duration (Rammsayer & Ulrich, 2001). Moreover, the duration estimation is based on the number of perceived changing events (Brown,1995). The speeds of the four progress bars changed continuously, whereas the constant progress bars remained the same. Hence, more changing events are detected by humans when seeing progress bars at a nonconstant speed. As a result, progress bars at a non-constant speed were perceived to be longer.

While the constant progress bar is perceived significantly faster than the slow-down, Bezier, and elasticity progress bars, there is no difference between the constant progress bar and speed-up.

A possible reason for the speed-up progress bar having better performance than the other three progress bars is anchoring. When people watch a progress bar move in real life, they tend to pay attention to the final period of the progress bar because it represents the end of the waiting process, which is a crucial timestamp for users to detect that the loading is nearly finished. As users would pay more attention to the end position of the visual feedback when waiting(Yokoyama et al., 2021), the final period of the progress bar may account for the result; within the last second of the progress bar movement, the velocity order is speed-up, elasticity, Bezier, and slow-down (Figure 3), which is precisely the reverse order of the PSE of the progress bars. This implies that when the final speed of the progress bar is faster, the progress bar is perceived faster. In other words, people may use final speed as an anchor to infer the time they have waited.

The anchoring effect occurs when people consider one number (an anchor), and their subsequent judgments are assimilated into it (Lee & Morewedg, 2022). Estimates are made by starting from an initial value that is adjusted to yield the final answer. As a result, different starting points yield different estimates that are biased toward the initial values (Tversky & Kahneman, 1974). For example, when people estimate the product as $8 \times 7 \times 6 \times 5 \times 4 \times 4 \times 2 \times 1$, the outcome is much greater than the estimation of $1 \times 2 \times 3 \times 4 \times 5 \times 6 \times 7 \times 8$.

In our study, to estimate the time duration of a non-constant progress bar, people may use the speed of the final period of the progress bar to adjust their estimation. Furthermore, if a progress bar's final speed is higher than another's, the insufficient adjustments make people feel that the progress bar with a higher final speed lasts shorter. Although the constant speed progress bar does not obey the anchoring effect, probably because people detect that the progress bar is constant before the final period, they would not use the speed to adjust their perception of time.

**Figure 3**

*The Speed of Different Progress Bars*

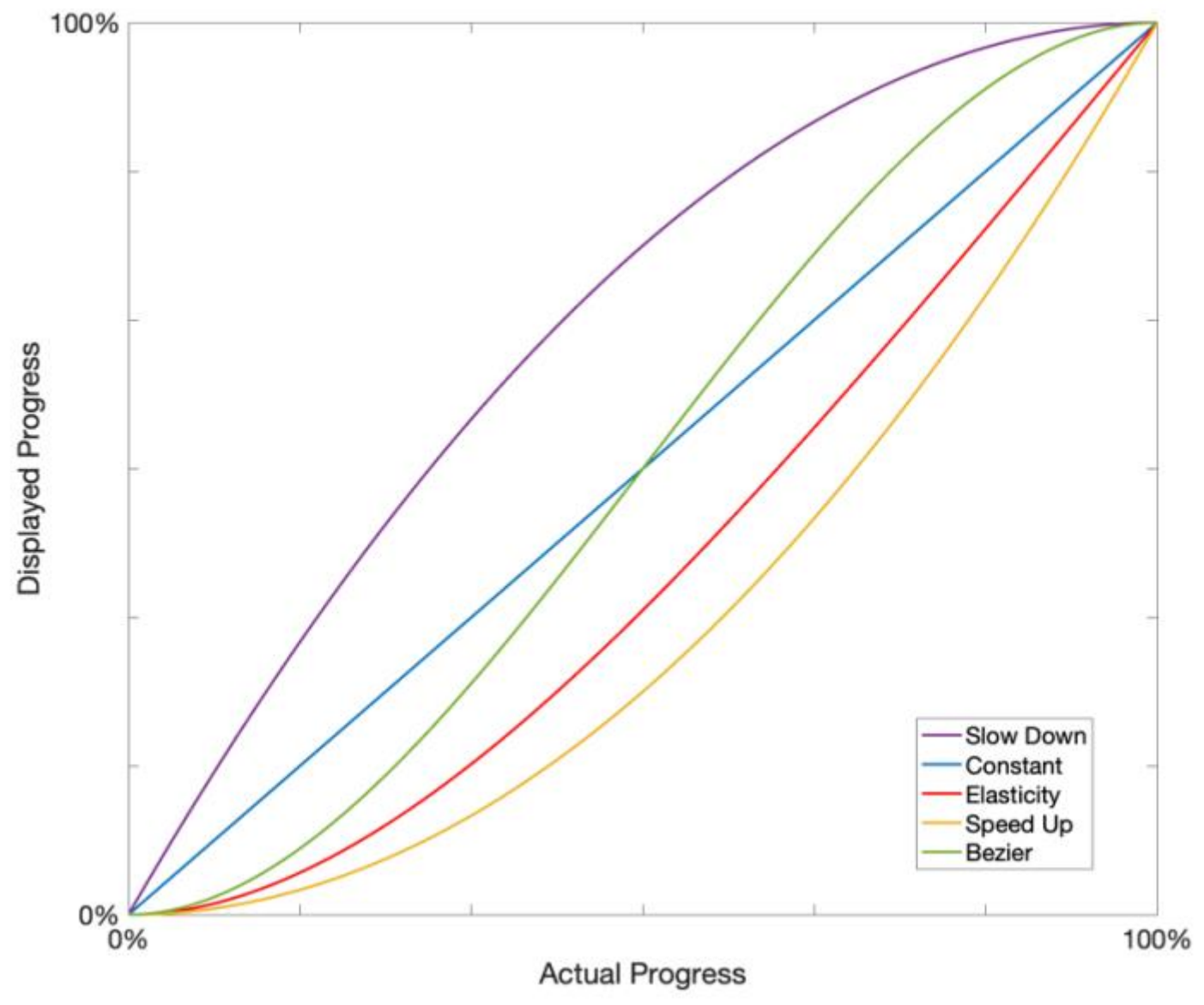

The eye-tracking results showed a strong negative correlation between the PSE of the progress bar and the EBR while watching the progress bar. Hence, the faster the progress bar is perceived, the larger the EBR. This can be explained in two ways.

First, the results indicate a relationship between cognitive load and PSE of progress bars. Some studies have shown that when participants participate in a cognitive task, the EBR is influenced by cognitive loads, defined as a multidimensional construct representing the amount of attentional and cognitive resources needed to perform a task correctly (Magliacano et al., 2020). The greater the cognitive load, the lower the EBR (Cardona et al., 2011). This relationship has been observed in many visual tasks, such as playing computer games (Cardona et al., 2011) and simulated air traffic control tasks (McIntire et al., 2014). On this basis, our findings support the idea that participants need to use higher cognitive resources to measure the time duration of the progress bar with greater PSE. In other words, the progress bar is perceived faster when the cognitive resource demand for the precepting time duration is lower. When people watch the progress bar with greater PSE, a considerable amount of visual attentional resources is required; consequently, they implement a semi-automatic strategy for reducing blinks to balance the current EBR between the needs for eye lubrication and those for maintaining efficient visual perception (Nakano et al., 2009). As the cognitive load increases, it is harder for humans to perceive the time duration of the progress bar, leading them to think that the duration is longer. This explanation is plausible for our result, as the progress bar with constant speed requires the least cognitive resources to measure the time duration compared to other progress bars. The cognitive load consists of limited working memory and vast long-term memory (Miller, 1956). A high cognitive load leads to an overload of working memory, resulting in a deterioration of perception (Sweller, 2005). The progress bar with greater PSE has a higher cognitive load; consequently, the working memory overload makes people feel that the time perception is longer.

Despite attentional resources, some studies also suggest that eye blinks are inhibited at precise moments to minimize the loss of visual information during a blink (Ranti et al., 2020). Thus, it is possible that people need more information to determine the time duration of progress bars with greater PSE. In conclusion, measuring the duration of PSE progress bars is more complex for human perception. When it is complex for humans to perceive, they tend to feel that the time duration is longer.

Our research is consistent with a previous study that suggested the advantages of a slow-to-fast progress bar (Harrison et al., 2007) and constant progress bar (Amer & Johnson, 2014) and is inconsistent with a previous study that suggested the benefits of the fast-to-slow progress bar (Conrad et al., 2010). This is because the progress bar and the methods we used are different. Conrad et al. (2010) used a survey to investigate users' time perception of progress bars, and the progress bar lasts for 5–7 mins since it is a progress bar to indicate the process of the task, not a loading progress bar. In contrast, our research and Harrison et al.'s (2007) research focus on short-time progress bars, using progress bars that last for 5 seconds and 5.5 seconds. Moreover, both our research and Harrison's research used forced-choice tasks repeated for many trials to allow participants to compare different progress bars directly, rather than using surveys, as in other studies (Conrad et al., 2010; Li & Chen, 2019). Other studies that support constant or fast-to-slow progress bars are better, perhaps because their methods lack validity, as they do not replicate the contrast many times; hence, some noise may obstruct the actual result. The differences in outcomes are due to differences in time and methods. Based on the fact that forced-choice tasks may reduce some noise in comparison, we admit that fast-to-slow progress bars may be perceived as shorter in long-term tasks, but slow-to-fast progress bars and constant progress bars are perceived to be shorter in short-loading processes, such as 5s. Consistent with this idea and in line with prior work, we strongly recommend using the slow-to-fast and constant progress bar in the UX design.

## Conclusion

Our experiment measured human time perception for different types of progress bars. We use the Bayesian model to conduct an adaptive experiment in which the constant progress bar duration is dynamically adjusted to be equal to four different non-constant progress bars, defined as the PSE. The results and explanations of our experiment are summarized as follows.

1. Our results show that all PSE are larger than the duration of the non-constant progress bars. The order of the non-constant progress bars from fastest to slowest is speed-up, elasticity, Bezier, and slow-down. Specifically, Bezier, elasticity, and slow-down progress bars were perceived to be significantly slower than constant progress bars. The slow-down progress bar is perceived to be significantly slower than the Bezier, elasticity, and speed-up progress bars.

2. The eye-tracking data show that the slower the progress bar is perceived, the lower the EBR.
3. This result can be attributed to the anchoring effect. To estimate the time duration of a nonconstant progress bar, people may use the speed of the final period of the progress bar to adjust their estimation.
4. Participants needed to use higher cognitive resources to measure the duration of the progress bar with a greater PSE. A higher cognitive demand causes users to think that the time duration is longer.

Using the Bayesian model to produce adaptive stimuli and with multiple trials, our study provides precise and accurate measurements of people's time perception, indicating that the constant progress bar and speed-up progress bar are perceived as the shortest. This may be useful for UX design and human-computer interactions.

**Limitation**

Our research only uses progress bars that last for 5s as a standard stimulus; whether the time perception for different progress bars changes with duration is still unclear. Future studies may consider time as a variable. Moreover, limited by the strict demand for psychophysics, the experiment was not presented in a real application scene, such as a video loading scene.

## Key Points

The result shows that the constant and speed-up progress bars are perceived faster than other progress bars.

Eye tracking shows that the slower the progress bar is perceived, the lower the EBR.

To estimate the time duration of a nonconstant progress bar, people may use the speed of the final period of the progress bar to adjust their estimation.

A higher cognitive demand causes users to think that the time duration is longer.

## References

Amer, T. S., & Johnson, T. L. (2014). IT progress indicators: sense of progress, subjective sense of time, user preference and the perception of process duration. *International Journal of Technology and Human Interaction (IJTHI)*, 10(3), 58-71.

Branaghan, R. J., & Sanchez, C. A. (2009). Feedback preferences and impressions of waiting. *Human factors*, 51(4), 528-538.

Brown, S. W. (1995). Time, change, and motion: The effects of stimulus movement on temporal perception. *Perception & Psychophysics, 57*(1), 105–116.

Cardona, G., García, C., Serés, C., Vilaseca, M., & Gispets, J. (2011). Blink rate, blink amplitude, and tear film integrity during dynamic visual display terminal tasks. *Current eye research*, *36*(3), 190-197.

Conrad, F. G., Couper, M. P., Tourangeau, R., & Peytchev, A. (2010). The impact of progress indicators on task completion. *Interacting with computers*, 22(5), 417–427.

Dang, J., Xiao, S., Liu, Y., Jiang, Y., & Mao, L. (2016). Individual differences in dopamine level modulate the ego depletion effect. *International Journal of Psychophysiology*, *99*, 121-124.

Eagleman, D. M., & Pariyadath, V. (2009). Is subjective duration a signature of coding efficiency?. *Philosophical Transactions of the Royal Society B: Biological Sciences, 364*(1525), 1841-1851.

Green, D. M., & Swets, J. A. (1966). *Signal detection theory and psychophysics* (Vol. 1, pp. 1969-2012). New York: Wiley.

Harrison, C., Amento, B., Kuznetsov, S., & Bell, R. (2007, October). Rethinking the progress bar. In *Proceedings of the 20th annual ACM symposium on User interface software and technology* (pp. 115-118).

Hornik, J. (1984). Subjective vs. objective time measures: A note on the perception of time in consumer behavior. *Journal of consumer research*, 11(1), 615-618.

Jongkees, B. J., & Colzato, L. S. (2016). Spontaneous eye blink rate as predictor of dopamine-related cognitive function—A review. *Neuroscience & Biobehavioral Reviews*, *71*, 58-82.

Karşılar, H., Kısa, Y. D., & Balcı, F. (2018). Dilation and constriction of subjective time based on observed walking speed. *Frontiers in psychology*, *9*, 2565.

Lee, C. Y., & Morewedge, C. K. (2022). Noise Increases Anchoring Effects. *Psychological Science*, *33*(1), 60-75.

Li, S., & Chen, C. H. (2019, July). The effect of progress indicator speeds on users' time perceptions and experience of a smartphone user interface. In *International Conference on Human-Computer Interaction* (pp. 28-36). Springer, Cham.

Magliacano, A., Fiorenza, S., Estraneo, A., & Trojano, L. (2020). Eye blink rate increases as a function of cognitive load during an auditory oddball paradigm. *Neuroscience Letters*, *736*, 135293.

Matthews, W. J. (2011). How do changes in speed affect the perception of duration?. *Journal of Experimental Psychology: Human Perception and Performance*, *37*(5), 1617.

McIntire, L. K., McKinley, R. A., Goodyear, C., & McIntire, J. P. (2014). Detection of vigilance performance using eye blinks. *Applied ergonomics*, *45*(2), 354-362.

Miller, G. A. (1956). The magical number seven, plus or minus two: Some limits on our capacity for processing information. *Psychological review*, *63*(2), 81.

Nakano, T., Yamamoto, Y., Kitajo, K., Takahashi, T., & Kitazawa, S. (2009). Synchronization of spontaneous eyeblinks while viewing video stories. *Proceedings of the Royal Society B: Biological Sciences*, *276*(1673), 3635-3644.

Rammsayer, T., & Ulrich, R. (2001). Counting models of temporal discrimination. *Psychonomic Bulletin & Review*, *8*(2), 270-277.

Ranti, C., Jones, W., Klin, A., & Shultz, S. (2020). Blink rate patterns provide a reliable measure of individual engagement with scene content. *Scientific reports*, *10*(1), 1-10.

Roberts, A. R., & Fishbach, A. (2022). Can't wait or won't wait? the two barriers to patient decisions. *Trends in Cognitive Sciences*, 26(4), 283-285.

Rose, G., Wu, W., & Yu, Y. (2019). Does subculture matter? A cross-cultural study of chronism and attitudes toward download delay in Internet systems in China and the United States. *Journal of Global Information Technology Management*, *22*(2), 82-99.

Simpson, W. A. (1989). The step method: A new adaptive psychophysical procedure. *Perception & Psychophysics*, *45*(6), 572-576.

Sweller, J. (2005). Implications of cognitive load theory for multimedia learning. *The Cambridge handbook of multimedia learning*, *3*(2), 19-30.

Taesler, P., & Rose, M. (2017). Psychophysically-anchored, robust thresholding in studying pain-related lateralization of oscillatory prestimulus activity. *JoVE (Journal of Visualized Experiments)*, (119), e55228.

Thomas, E. A., & Cantor, N. E. (1976). Simultaneous time and size perception. *Perception & Psychophysics*, *19*(4), 353-360.

Tversky, A., & Kahneman, D. (1974). Judgment under Uncertainty: Heuristics and Biases: Biases in judgments reveal some heuristics of thinking under uncertainty. *science*, *185*(4157), 1124-1131.

Vidotto, G., Anselmi, P., & Robusto, E. (2019). New Perspectives in Computing the Point of Subjective Equality Using Rasch Models. *Frontiers in Psychology*, *10*, 2793.

Villar, A., Callegaro, M., & Yang, Y. (2013). Where Am I? A Meta-Analysis of Experiments on the Effects of Progress. Indicators for Web Surveys. *Social Science Computer Review*, 31(6), 744–762.

Wang X., Wöllner C. (2019). Time as the ink that music is written with : A review of internal clock models and their explanatory power in audiovisual perception. *Jahrbuch Musikpsychologie* 29, 1–22.

Watson, A. B., & Pelli, D. G. (1983). QUEST: A Bayesian adaptive psychometric method. *Perception & psychophysics*, *33*(2), 113-120.

Yokoyama, K., Nakamura, S., & Yamanaka, S. (2021, August). Do Animation Direction and Position of Progress Bar Affect Selections?. *In IFIP Conference on Human-Computer Interaction* (pp. 395-399). Springer, Cham.

Zhang, T., Mou, D., Wang, C., Tan, F., Jiang, Y., Lijun, Z., & Li, H. (2015). Dopamine and executive function: Increased spontaneous eye blink rates correlate with better set-shifting and inhibition, but poorer updating. *International Journal of Psychophysiology*, *96*(3), 155-161.

**Biographies**

Qihan Wang

Yuanpei College, School of Psychological and Cognitive Sciences, Peking University

Peking University, Haidian District, Beijing, China

Phone: +86 15811097780

Email: wangqihan@pku.edu.cn

Qihan Wang is an Undergraduate student at Peking University. Research Interest: Human Computer Interaction, Social Computing.

Xinyue Kang

Dept. of Industrial Engineering, Tsinghua University

524A, Shunde Building, Tsinghua University, Haidian District, Beijing, China

Phone: +86 18711882988

Email: kxy20@mails.tsinghua.edu.cn

Xinyue Kang received her B.S. degree in Industrial Engineering from Tsinghua University, China, in 2020. She is a doctoral candidate at Tsinghua University. Her research interests include human-robot interaction, human-computer interaction and cross-culture design.

Pei-Luen Patrick Rau

Dept. of Industrial Engineering, Tsinghua University

525, Shunde Building, Tsinghua University, Haidian District, Beijing, China

Phone: +86- 010-6277-6664

Fax: +86- 010-6279-4399

Email: rpl@mail.tsinghua.edu.cn

Pei-Luen Patrick Rau received the Ph.D. degree in industrial engineering from Perdue University, USA. He is a professor of Industrial Engineering and vice dean of Global Innovation Exchange Institute at Tsinghua University. He is interested in the research areas of human factors engineering, human-computer interaction, cross-cultural design, and user experience.